# Comments on 'Growth and characterization of organic nonlinear optical crystal: *L*-Valinium salicylate (LVS)' [Optik 125 (2014) 2713-2715]

S. Natarajan[a], K. Moovendaran[a] and B. R. Srinivasan[b]
[a]School of Physics, Madurai Kamaraj University, Madurai 625 021, India
[b]Department of Chemistry, Goa University, Goa 403 206, India.

**Abstract**

The slow evaporation of an aqueous solution containing *L*-valine and salicylic acid results in the fractional crystallization of salicylic acid and not any so called '*L*-valinium salicylate' as reported recently by Andal and Murugakoothan in the title paper.

**Introduction**

 *L*-Valine, one of the twenty naturally occurring amino acids, is a component of several proteins, needed for muscle metabolism, repair and growth of tissue and maintaining the nitrogen balance in the body. The crystal structure of *L*-valine [1] and several of its salts are well documented in literature [1-7]. As part of a long standing research program on the synthesis, structure elucidation and bulk growth of amino acid-based NLO crystals [8-9], the growth of *L*-valine crystals and the crystal structures of several compounds of *DL*-valine [10-16] have been reported from our laboratory. Our studies in this area have revealed that *L*-valine does not prefer to form *L*-valinium salts with carboxylic acids [15-16] unlike its well documented behavior of formation of *L*-valinium salts with strong mineral acids such as $HNO_3$, $HClO_4$, $H_3PO_4$, etc. Our efforts to crystallize new compounds of *L*- or *DL*-valine using aromatic acids like benzoic or salicylic acid were not fruitful. Hence the title paper [17] reporting the growth of *L*-valinium salicylate (LVS) crystals by slow evaporation method attracted our attention. A scrutiny of the paper revealed that LVS was not characterized based on a refinement of the X-ray structure despite the claim of single crystal work. The reported unit cell data (*a*, *b*, *c* parameters) perfectly matched with those for salicylic acid reported by Sundaralingam and Jensen in 1965 [18] in the monoclinic space group *P2₁/a*. However, the authors of the title paper reported a triclinic structure. It appeared that the characterization of LVS was improper. In view of the absence of a CIF file to prove the structural refinement of LVS and also our interest in compounds of *L*-valine, we have reinvestigated the crystal growth reported in the title paper. The results of these investigations are described in this comment.

**Reinvestigation of the crystal growth of *L*-valinium salicylate**

Commercially available *L*-valine (Loba), salicylic acid (Sisco Research Laboratories) and water and methanol were used for the crystal growth. A mixture of *L*-valine (1.1715 g) and salicylic acid (1.3812 g) were taken in the stochiometric ratio, dissolved in a mixed solvent of water and methanol (1:1) and stirred well to obtain a clear solution. Then the solution was filtered and the clear filtrate left undisturbed for crystallization. Slow evaporation of solvent at ambient temperature (30 ºC) resulted in the separation of transparent crystals after 4-5 days. The crystals thus obtained were labeled as compound **1** and characterized by its melting point, infrared (IR) spectrum and X-ray powder diffractogram and these data were compared with those of salicylic acid (**2**) recrystallized from methanol. Melting points were measured using a melting point apparatus. IR spectra of the samples were recorded in KBr matrix in the frequency region of 400 to 4000 cm$^{-1}$, using a Jasco spectrometer (FTIR, model 410) with a resolution of 4 cm$^{-1}$. Powder X-ray diffractograms were recorded using a PANalytical X'Pert Pro (DY1850) powder diffractometer with Cu Kα radiation (λ = 1.54056 Å). The sample was scanned using a step size of 0.01°/sec in the 2θ range of 10–60°.

**Comment**

The crystals (**1**) grown in this work under reported conditions in the title paper melted at 158(1) °C which is in good agreement with the melting point of **2** 159(1) °C showing **1** is pure salicylic acid. This was unambiguously confirmed by a comparison of the IR spectrum and X-ray powder diffractogram of **1** with those of pure salicylic acid (Fig. 1). The perfect matching of the IR spectra and the powder diffractograms proves that compound **1** is pure salicylic acid.

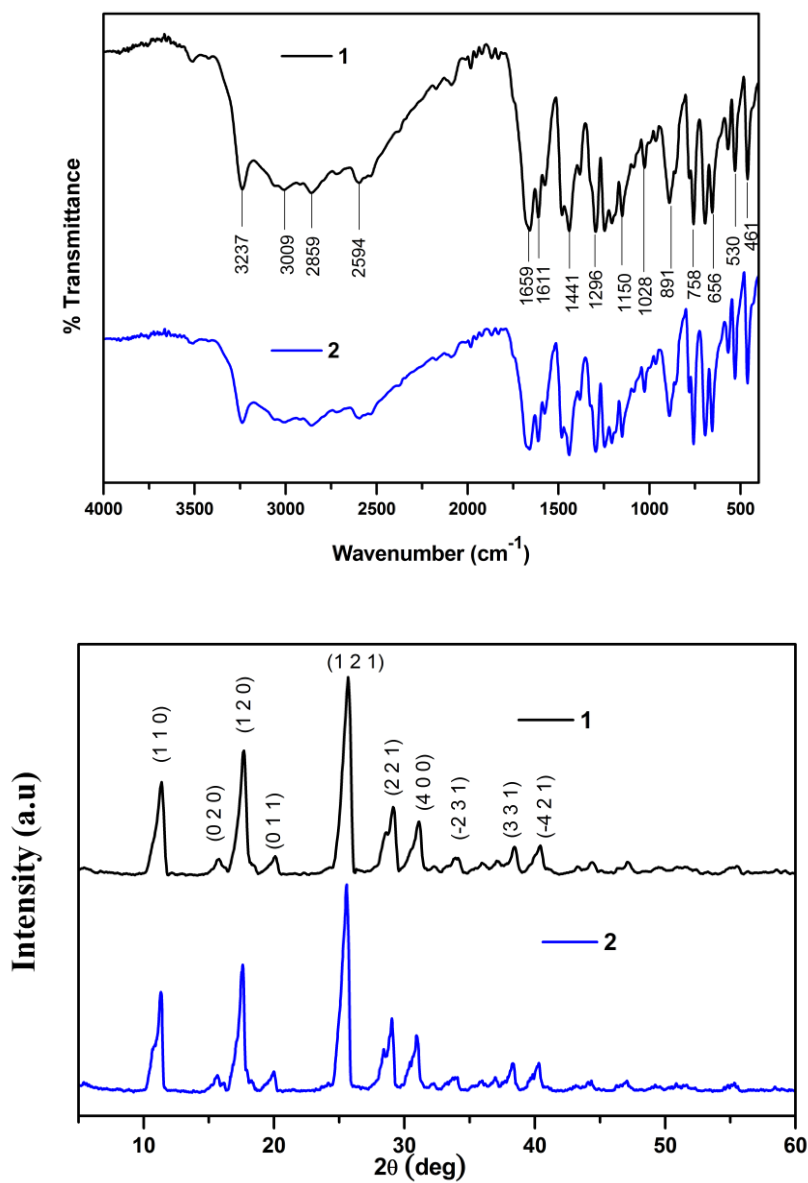

Fig. 1. IR spectra (top) and X-ray powder patterns (bottom) showing that a so called *L*-valinium salicylate (**1**) (black line) and salicylic acid (**2**) (blue line) are one and the same.

As the exact identity of the product crystal from the slow evaporation solution growth of *L*-valine and salicylic acid was correctly established as salicylic acid based on melting point, IR spectrum and X-ray powder diffractogram, we considered the determination of single crystal structure a superfluous exercise. In this context it is to be noted that the unit cell parameters (Table 1) reported for **1** in the title paper are in very good agreement with literature data for salicylic acid. However, it is not clear as to how the authors could classify their crystal as triclinic without any structure refinement.

Table 1. Reported X-ray data on salicylic acid

| Compound name | $a$ (Å) $\alpha$(°) | $b$ (Å) $\beta$ (°) | $c$ (Å) $\gamma$(°) | Space Group | Ref |
|---|---|---|---|---|---|
| Salicylic acid | 11.52(2) | 11.21(1) 90.50(2) | 4.92(5) | $P2_1/a$ | [18] |
| *L*-valinium salicylate | 11.53# 90.02# | 11.20# 90.65# | 4.91# 89.96# | * | [17] |

Note – # No *e.s.d.* values reported; * No space group reported

The authors of the title paper have formulated their LVS crystal based on an incorrect assumption that slow evaporation of a solution containing equimolar quantities of *L*-valine and salicylic acid will result in the formation of LVS. The formation of salicylic acid as the product crystal can be explained due to no chemical reaction between *L*-valine and salicylic acid and the fractional crystallization of the less soluble reactant salicylic acid [19], with the more soluble *L*-valine remaining in solution. It is unfortunate that the authors did not take into consideration the chemistry of the reagents as can be evidenced by the fact that none of the citations in the title paper were for any *L*-valine or salicylic acid based compound.

It is not clear as to how the authors of the commented paper could measure an SHG response of three times that of KDP for their so called LVS which is actually salicylic acid and claim, '*It is very clear that this crystal has potential and prospects for frequency conversion applications*'. Since salicylic acid is a centrosymmetric solid, the only possible explanation is that no such measurement was probably made. Several such claims of growing novel NLO crystals based on unit cell data or observation of strong SHG signals in centrosymmetric solids have been proved to be erroneous by the groups of Fleck, Petrosyan, Srinivasan and Natarajan [20-27] and *L*-valinium salicylate is another dubious NLO crystal. Our present results once again show the risk of formulating new compounds based only on unit cell data, a single IR spectrum or an NMR spectrum instead of structure refinement based on single crystal data.

**Conclusions**

The main findings of the present work are the following: (i) It is shown that the reported 'L-valinium salicylate'[17] is not a novel non-linear optical single crystal, but instead the well-known salicylic acid crystal, (ii) the measurement of melting point is a simple but useful technique to distinguish solid products from the reagents used, (iii) a comparison of the IR spectra and X-ray powder diffractograms offers a convenient method to distinguish the products of a reaction from the starting materials and (iv) it is incorrect to formulate a new compound based only on X-ray unit cell data and also to report space group without structure determination.